\documentclass[12pt]{iopart}
\begin{document}
\include{epsf}

\jl{3}      

\title{Wetting at Non-Planar and Heterogeneous Substrates}

\author{C Rasc\'{o}n and A O Parry}

\address{Mathematics Department, Imperial College, \\
180 Queen's Gate, London SW7 2BZ,  UK}

\begin{abstract}

We report results of wetting on non-planar and heterogeneous
surfaces calculated from an effective interfacial Hamiltonian
model. The lack of translational invariance along the
substrate induces a series of structural changes on the
interface such as unbending and a number of non-thermodynamic
singularities and can modify the location of the wetting
transition. We show that the order of the wetting
transition in the planar homogeneous system strongly affects
the behaviour of the non-planar and heterogeneous
surfaces. 
\end{abstract}




\section{Introduction}
The interaction of fluids with solid substrates is attracting
new interest as experimental methods allow increasing
control over the shape and chemical composition of solid surfaces
\cite{Surfaces}. The theoretical description of fluid adsorption
necesarily involves breaking
the fluid translational invariance due to the presence of a wall,
representing the solid substrate. Due to the intrinsic
difficulty of this, theoretical studies have concentrated mainly on
planar and homogeneous substrates producing a deep understanding
of the rich behaviour of those systems \cite{Dietrich}. However,
non-planar and chemically heterogenous surfaces exhibit adsorption
properties which differ from those of planar and homogeneous systems
and require further study \cite{Rough}. Here, we use an effective
interfacial Hamiltonian model to examine the wetting
properties of a corrugated substrate, and a planar substrate with a
stripe of a material chemically different.

\begin{figure}[ht]
\vspace*{-4cm}
\label{first}
\centerline{\epsfxsize=16cm \epsfbox{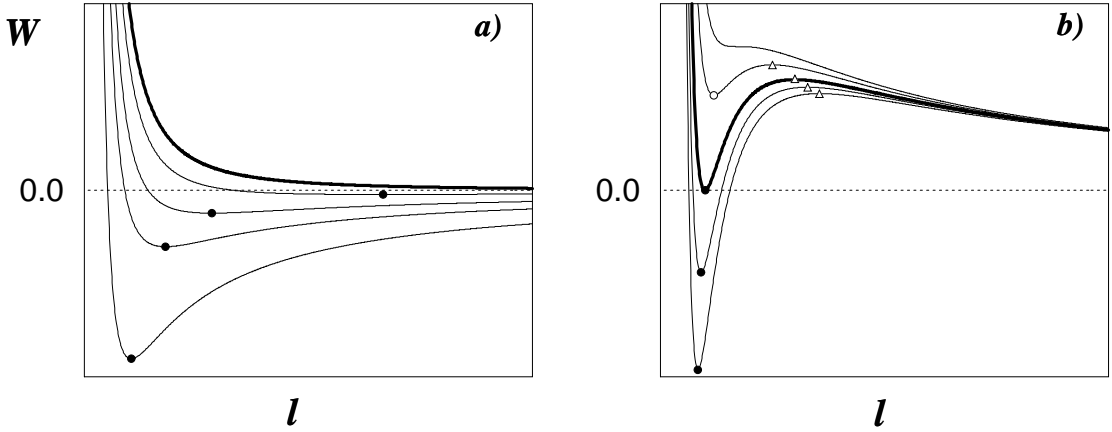}}
\vspace*{-3.cm}
\caption{Effective binding potentials for a) second-order and
b) first-order wetting transitions in planar homogeneous systems.
The minimum of the potential (\fullcircle) represents the
equilibrium configuration. At the wetting temperature (thick line),
that configuration has zero energy. Note the activation barrier
(\opentriangle) and the existence of metastable states (\opencircle)
in the first-order potential.}
\end{figure}

\section{The Model}

For simplicity, we only break the symmetry along one of the
directions of the wall and, therefore, one coordinate ($x$ say) will
describe any point on the surface. The free energy of an interfacial
configuration is given by the standard effective Hamiltonian
\cite{Dietrich}
\begin{equation}
\label{one}
H[\ell]=\int\!dx
\;\left[\,{\Sigma\over 2}\left({{d\ell}\over{dx}}\right)^2
+{\cal W}(\ell;x)\,\right]
\end{equation}
where $\ell(x)$ represents the height of the fluid interface,
$\Sigma$ is the interface stiffness and
${\cal W}(\ell;x)$ accounts for the (effective) interaction with the
substrate. We can now anticipate that the new phenomena occuring for
non-planar or heterogeneous systems take place due to the competition
of the two terms in the Hamiltonian; whilst the first term forces the
interface to minimize its extent, the second is constrained by the
intermolecular forces between the particles. The character of these
interactions is qualitatively captured by the following expression
for the potential ${\cal W}$,
\begin{equation}
\label{two}
{\cal W}(\ell;x)\;=\;W_{\gamma}\left(\,\ell-\psi(x)\,\right)
\hspace*{1.5cm}\hbox{for }x\in\Lambda_{\gamma}
\end{equation}
where $\psi(x)$ is the height of the wall at the point $x$ and
$W_{\gamma}$ is the effective binding potential of a fluid interface
on a {\it planar} and {\it homogeneous} substrate
(which extends along $\Lambda_{\gamma}$).
A planar wall corresponds to $\psi(x)=0$ whilst,
for a chemically homogeneous wall, there is only one region $\Lambda$
and consequently only one binding potential $W$. The approximation
(\ref{two}) is appropriate for walls whose non-planarity is not
too severe (see later for further quantification).

Although effective binding potentials are well described in the
literature \cite{Dietrich}, we want to outline some of their
features for a subsequent discussion.
In figure 1,
two different types are plotted for different values of the temperature
(at bulk liquid-vapour coexistence). For second-order wetting, (a), the
potential has a single minimum, located at $\ell\!=\!\ell_{\pi}$, for
$T<T_{W}$. The coverage of the planar system, $\ell_{\pi}$, diverges
at the wetting temperature $T_{W}$. In contrast,
for first-order wetting, (b), the potential shows a minimum at
$\ell\!=\!\ell_{\pi}$ and a maximum, the activation barrier, at
$\ell\!=\!\ell_{\star}$. In this
second case, the thickness of the adsorbed layer remains finite at
the wetting temperature, coexisting with an
infinitely thick layer, $W(\ell_{\pi}^{W})\!=\!W(\infty)\!=\!0$
(See figure 1). Furthermore, for a range of temperatures
$T_{W}\!<\!T\!<\!T_{S}$, where $T_{S}$ is the spinoidal temperature,
the first-order effective potential still shows a
minimum which represents a thin layer
metastable with respect to the infinitely thick one,
$W(\ell_{\pi})\!>\!W(\infty)\!=\!0$.

\section{Results}

We restrict ourselves to a mean-field description of the system and,
to calculate the equilibrium profile, we minimise the Hamiltonian (\ref{one}),
which is equivalent to solving the Euler-Lagrange equation:
\begin{equation}
\label{five}
{{d^{\,2}\ell}\over{dx^{2}}}\;=\;W'_{\gamma}
\left(\,\ell-\psi(x)\,\right)
\hspace*{1.4cm}\hbox{for }x\in\Lambda_{\gamma}.
\end{equation}
The results depend sensitively on the character of the wetting
transition on the planar substrate and, therefore, we present some
representative results for non-planar surfaces according to the order
of the wetting transition on the planar substrate.

\subsection{Non-Planar Walls}

\subsubsection{Second-order wetting binding potentials.}

First, we consider a homogeneous corrugated wall, with
$\psi(x)\!=\!a\cos(2\pi x/L)$, where $a$ represents the corrugation
amplitude and $L$ the period of the corrugation.
We assume that the wavelength $L\gg a$ so that the corrugation
is relatively weak. In this limit, scaling properties emerge
which are correctly captured by the assumption of a vertical
height interaction in the effective binding potential
(see (\ref{two})).
Details of this problem for a second-order binding potential
have been given elsewhere \cite{RPS} but we
report the results here for the sake of comparison. In this case,
the interface undergoes a first-order {\it unbending} transition
at a temperature $T$ {\it below} the wetting transition $T_{w}$
provided the corrugation exceeds a certain threshold (See figure 2 (left)).
At low temperatures,
the interface closely follows the corrugations so that both the
interface and wall,
have a similar shape. Above the transition temperature, however,
the interface is significantly flatter (unbent). The difference
between these coexisting states reduces with
the corrugation and it disappears at a critical point (\fullcircle). 
As pointed out, this transition takes place due to the
competition of the two terms of the Hamiltonian (\ref{one}).
Whilst the system minimises the functional at low temperatures
by following the shape of the surface (with a large negative
contribution of the second term of the Hamiltonian), the unbent
configuration reduces the energy by decreasing the (positive)
contribution of the first term. Interestingly,
the period of the corrugation does not change the structure of the
surface phase diagram but acts as a scaling parameter \cite{RPS}.
Within this model, neither the location (at $T\!=\!T_{w}$)
nor the character of the wetting (unbinding) transition is modified.

\begin{figure}[ht]
\label{second}
\centerline{\epsfxsize=8cm \epsfbox{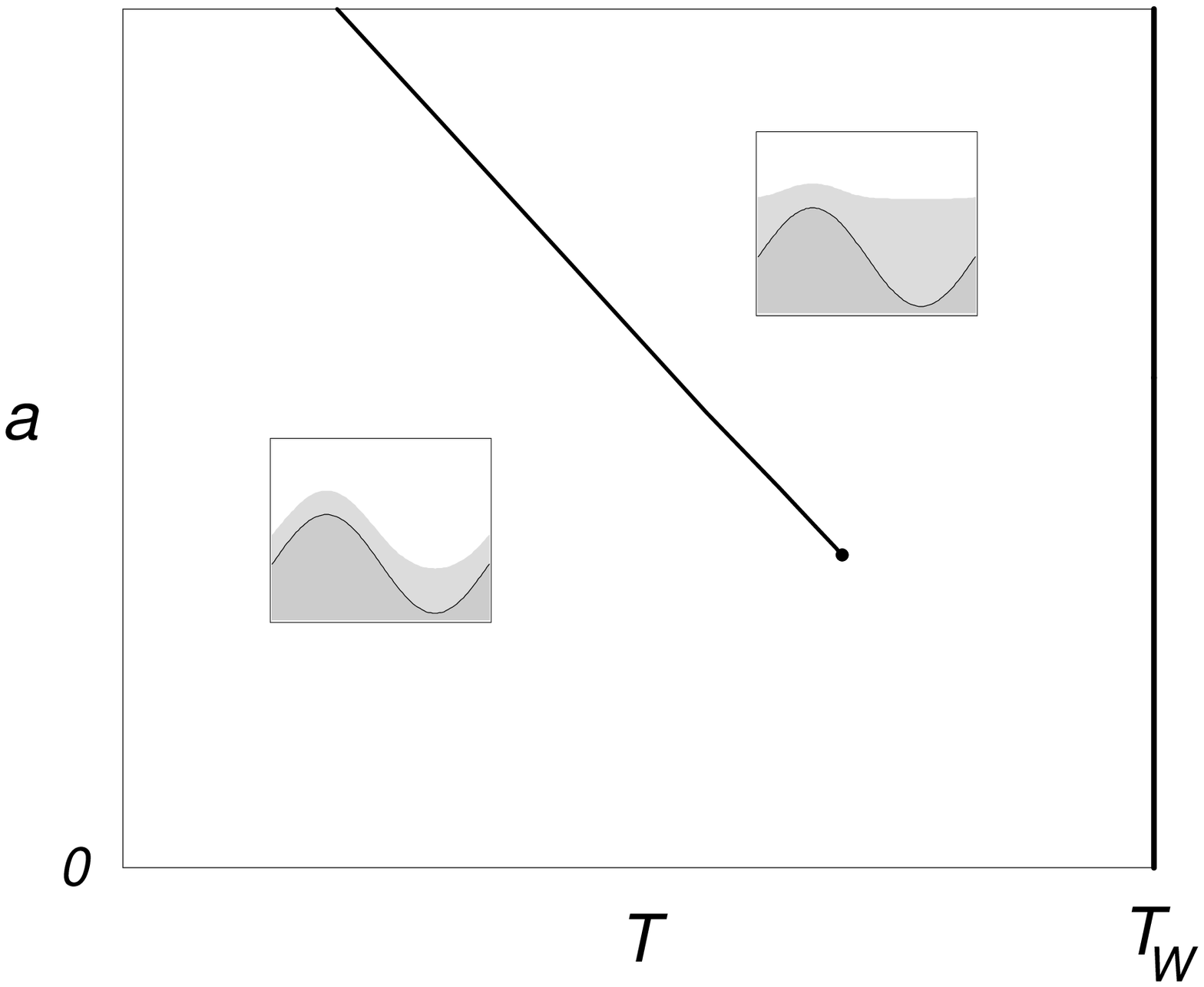}
\epsfxsize=8cm \epsfbox{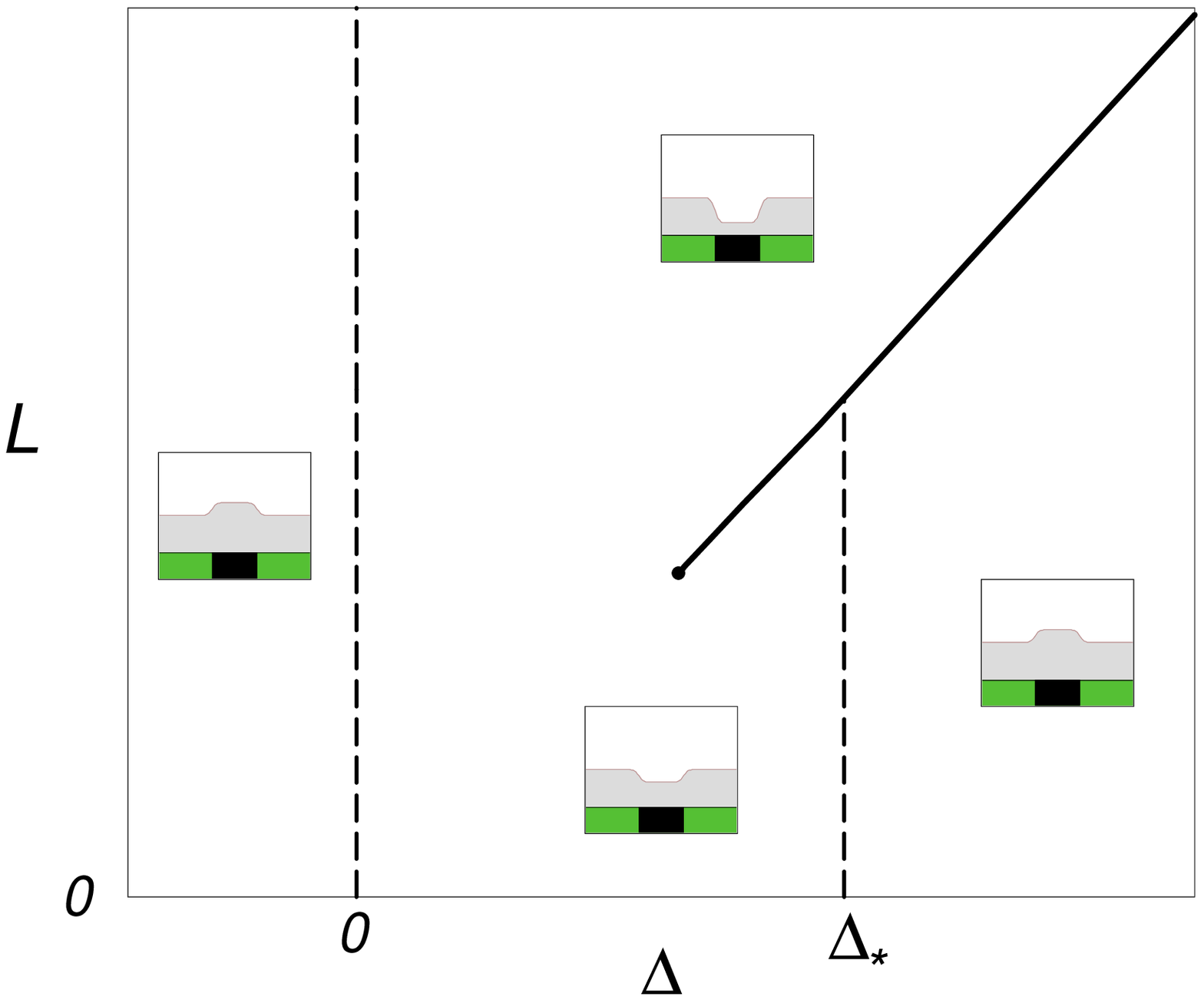}}
\caption{Schematic phase diagram of the structure of
an adsorbed layer of liquid on a homogeneous corrugated wall
(left) and on a planar substrate with a stripe of a material
chemically different (right). Bold lines represent expected
phase transitions. For the heterogeneous wall,
the dashed lines illustrate non-thermodynamic singularities
where the interface shows a qualitative change of shape in
which a trough becomes a crest and viceversa.}
\end{figure}

\subsubsection{First-order wetting binding potentials.}

If the binding potential is first-order, the surface phase
diagram is richer and the effect of the corrugation is triple.
First, an {\it unbending} transition can also
take place as the above mentioned competition between both terms
of the Hamiltonian is still present. However, first-order effective
potentials have an activation barrier which allows the
interface to adopt a variety of shapes to minimize the energy.
These shapes can be characterised by the number of minima of the
interface $\ell(x)$ and the change of this number gives rise to
a number of non-thermodynamic singularities. These singularities
have been studied for a corrugated wall in a related system
(in the context of confinement) \cite{RP}. Figure 3 (left) shows
schematically the possible configurations of the interface. Note
that, in this case, the interface shape can deviate significantly
from the wall shape and adopt configurations which are not found
with a second-order effective potential. The third effect of
the corrugation on this type of surface comes from the fact that
the energy of the unbound state ({\it i.e.}, the wet configuration,
$\ell\!=\!\infty$) is always zero. Therefore, the wettability
is favoured by any positive contribution to the Hamiltonian and
the wetting temperature is reduced. This effect is absent in
the second-order wetting potential because the interface
always finds a configuration whose positive contribution (first
term of the Hamiltonian) is lower than the negative (second term)
and the balance remains negative for any corrugation. The presence of
the activation barrier makes this compromise impossible and
the wetting transition takes place at a {\it lower} temperature
in the corrugated system than in the planar one, although the
nature of the transition is still first-order. Figure 3 (right) shows
a typical variation of the wetting temperature as a function of the
corrugation amplitude (in units of that of the planar system).
The periodicity of the corrugation, as
in the previous case, only acts as a scaling parameter (this is a
property of the Hamiltonian (\ref{one})). Note that the temperature
drops as a function of the corrugation amplitude, $a$, but it
presents a minima for $a\sim 1.5$ (in units of liquid bulk correlation-length).
For larger values of $a$, the
tendency is inverted and the wetting temperature increases slightly
and tends to a certain constant value. This behaviour can be traced
back to the varying shape the interface can adopt to minimize the
energy \cite{RP2}. As a result of this non-monotonic variation of
the wetting temperature with corrugation, the shape of the interface
at the wetting transition itself shows similar sensitivity \cite{RP2}.

\begin{figure}[ht]
\label{third}
\centerline{\epsfxsize=8cm \epsfbox{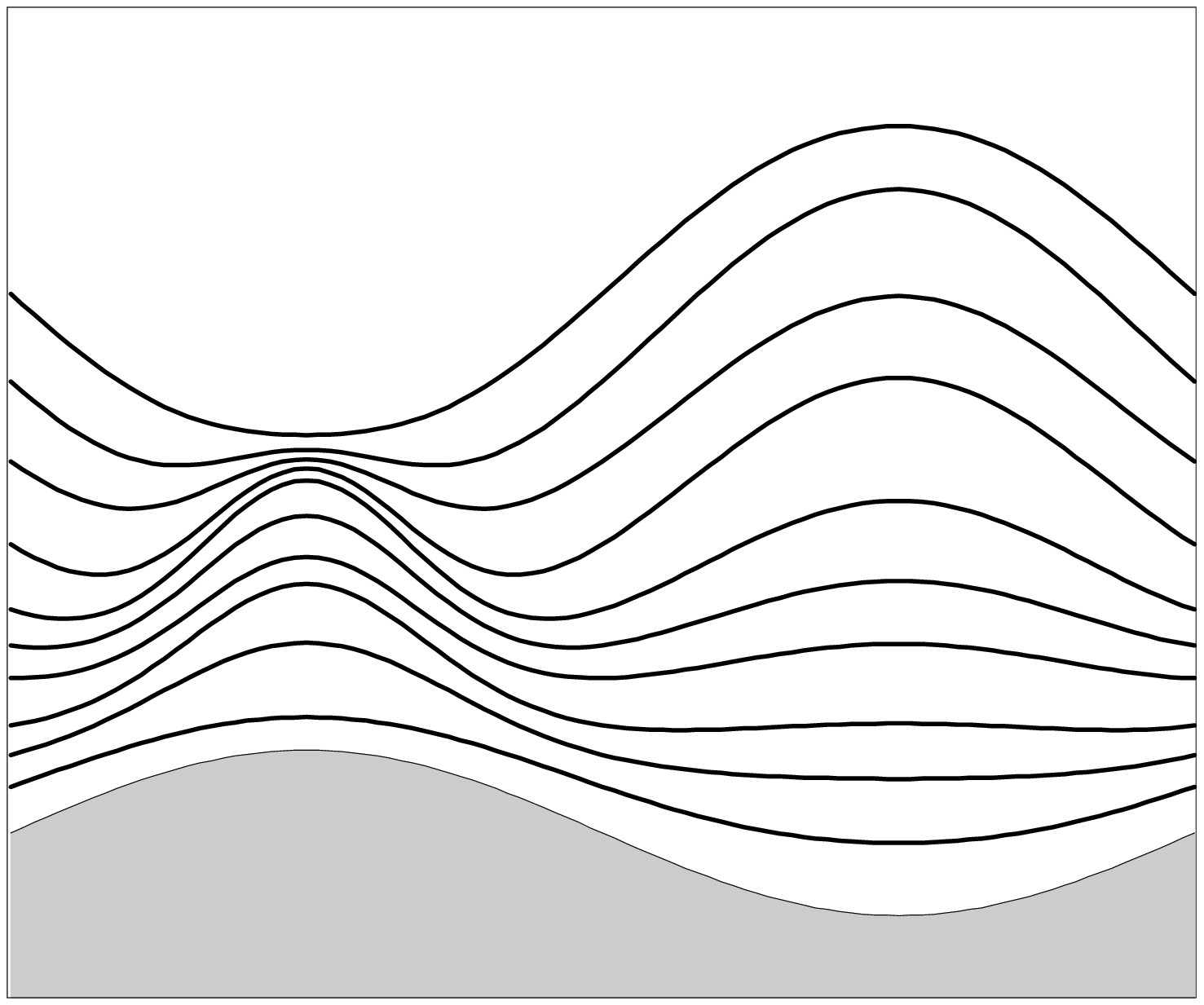}
\epsfxsize=8cm \epsfbox{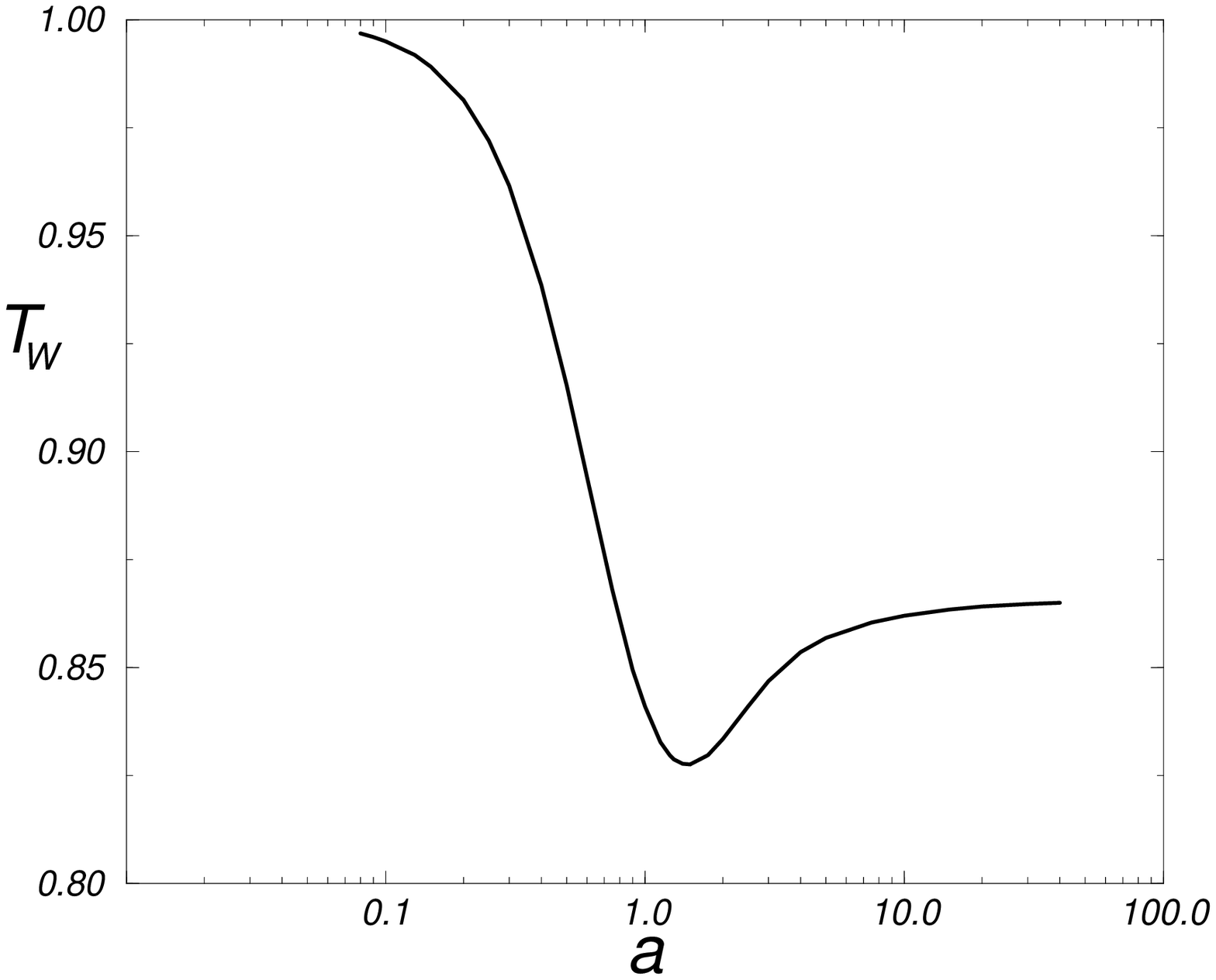}}
\caption{Effect of corrugation on the wetting properties of an
interface which (in the planar case) undergoes a first-order
wetting transition: the left hand diagram illustrates schematically
how the interface moves away from the wall, on increasing the
temperature (at fixed corrugation), and adopts a number of
distinct shapes which differ from that of the wall (lowest line).
The wetting temperature is also reduced by corrugation (right).
See text for details.}
\end{figure}

\subsection{Heterogeneous Walls}

We now focus our attention on (planar) heterogeneous walls. As
mentioned, we study an infinite homogeneous and planar
substrate (called 2) with a stripe of constant width $L$ of a chemically
different material (called 1), represented by two different
effective potentials. This geometry allows us to concentrate
only on the structural changes of the interface due to the
heterogeneity since the infinite substrate 2 governs the wetting
behaviour of the whole system \cite{Bauer}. The influence of heterogeneity
on the wetting properties of a substrate (for instance, due to
a periodic array of stripes) is a more complex problem which
requires the prior understanding of this simpler system.
Without loss of generality, we can
consider that homogeneous substrates 1 and 2 undergo first and
second-order wetting transitions respectively. The
structure of the phase diagram of this system does not depend on
the order of the wetting transition of the infinite system.
In fact it depends mainly on two quantities only: the stripe width, $L$,
and the {\it mismatch} between the effective potentials of
both substrates (at fixed temperature), {\it i.e.},
the difference between the thickness of the adsorbed layers in
the infinitely homogeneous systems,
$\Delta\!\equiv\!\ell_{\pi}^{(2)}-\ell_{\pi}^{(1)}$.
Figure 2 (right) shows a schematic phase diagram as a funcion of these
variables. As expected, the interface is flat if the mismatch
is zero ($\Delta\!=\!0$,\longbroken). For small values of $\Delta$, positive
or negative, the thickness of the interface above
the heterogeneity roughly behaves as in the infinite system.
For narrow stripes ($L\!\sim\!0$), however, we found that the
interface also flattens when $\Delta$ reaches the value
$\Delta_{\star}\!\equiv\!\ell_{\star}^{(1)}-\ell_{\pi}^{(1)}$
(\longbroken). At that point, the minimum
of potential 2 matches the {\it maximum} of potential
1, $\ell_{\pi}^{(2)}\!=\!\ell_{\star}^{(1)}$. Surprisingly, the
flat configuration is stable even though it would be unstable
for an infinite system. Nevertheless, if the stripe width exceeds a
certain value, the flat configuration becomes metastable with respect to
a non-flat one. This is due to the crossing of a line of 
generalized {\it unbending} transitions (\full)
which arises from the ubiquitous balance between the
two terms in Hamiltonian (\ref{one}). This line ends at a critical
point $(\Delta_{c},L_{c})$ (\fullcircle). Note that the configurations
along the coexistence line correspond to those of an unbending transition as
mentioned above if $\Delta_{c}<\Delta<\Delta_{\star}$ but, for
$\Delta>\Delta_{\star}$, the coexisting states are interfaces bent
in opposite directions. This second part of the line has its origin
in the existence of an activation barrier of the effective potential
of the heterogeneity. At this point, we note that the structure of
the phase diagram when the stripe effective potential is second
order (no activation barrier) can be intuitively obtained
from that in figure 2 (right) by considering $\Delta_{\star}\rightarrow\infty$,
thus recovering a usual unbending transition.

As a last remark, we want to discuss the twofold
role of the temperature in this description.
On the one hand, the phase diagram deforms (although
the general features are conserved). On the other, the value
of $\Delta$ varies. We can anticipate that from these two
competing tendencies the phase diagram can show a complex
behaviour including reentrant phases \cite{RP2}.

\ack  CR acknowledges economical support from the European Commission
under contract ERBFMBICT983229.


\Bibliography{1}

\bibitem{Surfaces} See, for example, Rockford L \etal 1999 \PRL {\bf 82} 2602
\nonum Trau M \etal 1997 Nature {\bf 390} 674
\nonum Xia Y and Whitesides GM 1996 Adv.\ Matter.\ {\bf 8}, 765
\bibitem{Dietrich} For a review, Dietrich S 1988 in {\it Phase Transitions and
Critical Phenomena} vol~12 ed C Domb and JL Lebowitz (London, Academic Press)
p~1
\bibitem{Rough} For a recent review, Dietrich S 1998 in Proc.\ of the
NATO-ASI Conf.\ {\it New Approaches to Old and New Problems in Liquid State
Theory}  ed C Caccamo, JP Hansen, G Stell
\bibitem{RPS} Rasc\'{o}n C, Parry AO and Sartori A 1999 \PR E {\bf 59} 5697
\bibitem{RP} Rasc\'{o}n C and Parry AO 1998 \PRL, {\bf 81} 1267
\bibitem{RP2} Rasc\'{o}n C and Parry AO 1999 (in preparation)
\bibitem{Bauer} C Bauer and S Dietrich ({\tt cond-mat/9906168})

\endbib
\end{document}